\documentclass[
   final           % use final for the camera ready runs
  ]
  {aipproc}
\usepackage{graphicx}
\usepackage{url}
\usepackage{amssymb,amsmath,epstopdf}

\layoutstyle{6x9}

\begin{document}

\title{Transverse Beam Size Effects on Longitudinal Profile Reconstruction}


\author{G. Andonian}{
  address={University of California Los Angeles, Particle Beam Physics Laboratory}
}
\author{E. Hemsing}{
  address={University of California Los Angeles, Particle Beam Physics Laboratory}
}
\author{A. Murokh}{
 address={RadiaBeam Technologies, LLC}
} 
\author{M. Dunning}{
  address={University of California Los Angeles, Particle Beam Physics Laboratory}
}
\author{G. Marcus}{
  address={University of California Los Angeles, Particle Beam Physics Laboratory}
}
\author{J. Rosenzweig}{
  address={University of California Los Angeles, Particle Beam Physics Laboratory}
}
\author{O. Williams}{
  address={University of California Los Angeles, Particle Beam Physics Laboratory}
}
 \author{V. Yakimenko}{
 address={Brookhaven National Laboratory, Accelerator Test Facility}
 }

\begin{abstract}
The use of coherent transition radiation autocorrelation methods to determine bunch length and profile information is examined with the compressed electron beam at the BNL ATF. 
A bi-gaussian fit
is applied to coherent transition radiation auto-correlation data to extract the longitudinal current distribution. 
The effects of large transverse beam sizes are studied in theory and compared to experimental results. 
A suitable form of the correction factor is derived for beams with large transverse-longitudinal aspect ratios.
\end{abstract}

\maketitle
%%%%%%%%%%%%%%%%%%%%%%%%%%%%%%%%%%%%%%

\section{Introduction}
Progress towards short wavelength free-electron laser (FEL) necessitates the production of high average current electron beams \cite{LCLS}. The required high-average currents for the x-ray FEL are typically attained using bunch compression techniques such as compression with a magnetic chicane (four dipole array) or velocity bunching techniques \cite{compression}. 
The measurement of the bunch length is crucial for reliable operation and delivery of consistent beam quality.

Several methods are commonly employed to extract the bunch length of ultra short electron beams
such as rf deflection, zero-phasing, electro-optic sampling, and spectral reconstruction methods 
\cite{GeitzDIPAC1999,lai}. Spectral methods use coherent radiation emitted from the electron beam that contains information about the longitudinal bunch profile.
Here, we examine features of a common spectral technique in which the bunch length is reconstructed from the measured spectrum of coherent transition radiation (CTR) emitted from an insertable foil \cite{HappekPRL1991}. The CTR signal is auto-correlated via interferometric methods, and the auto-correlation is fit to a simple bi-gaussian function to extract the pulse length. This technique is generally robust for simple beam geomerties, but requires a correction factor to account for high frequencies that are suppressed in the CTR spectrum when the transverse size is much larger than the bunch length (``pancake" beams). The missing frequencies result in a narrowed spectrum and thus an artificially lengthened bunch profile in the reconstruction. This paper describes a measurement carried out at the Brookhaven National Laboratory Accelerator Test Facility (BNL ATF) to investigate this effect. An analytic function is obtained that relates the reconstructed length to the actual length, which allows one to correct for the effect of the finite beam distribution.

\section{Transverse Beam Size Dependence}

\subsection{Analytical Approach}
The coherent, far-field angular spectral emission from an axi-symmetric gaussian electron beam striking a perfectly conducting surface is given by\cite{ShibataPRE501994}, 
 \begin{equation}
\frac{\mathrm{d}^2U_C}{\mathrm{d}k \mathrm{d}\Omega}=\frac{N_e^2e^2}{4\pi^3\epsilon_0}\frac{\chi(\theta)\sin^2{\theta}}{(1-\beta^2\cos^2{\theta})^2}e^{-k^2(\sigma_z^2\cos^2\theta+\sigma_r^2\sin^2\theta)},
\label{eq:didw}
\end{equation}
%\begin{equation}
%\frac{\mathrm{d}^2U_C}{\mathrm{d}k \mathrm{d}\Omega}=\frac{N_e^2e^2}{4\pi^3\epsilon_0}\chi(\theta)F(k)\frac{\sin^2{\theta}}{(1-\beta^2\cos^2{\theta})^2}
%\label{continuousfield}
%\end{equation}
where $U_C$ is the photon energy, $k=\omega c$ is the wavenumber, $N_e$ is the number of electrons, $\beta c$ is the longitudinal velocity (for this discussion, the beam is assumed relativistic so $\beta\simeq 1$), $\beta^2=1-\gamma^{-2}$, $\mathrm{d}\Omega=\sin\theta\mathrm{d}\theta\mathrm{d}\phi$ is the infinitesimal solid angle, $\theta$ is the forward opening angle, and $\phi$ is the azimuthal angle. $\chi(\theta)$ is the divergence factor that quantifies the contribution from angles in the beam, and is set to unity for the cold beam model considered here. 
%The finite e-beam distribution is given by the form factor $F(k)$ given by the spatial Fourier transform of the normalized e-beam charge distribution $f(\mathbf{x})=f_{\perp}(\mathbf{x}_{\perp})f_z(z)$: 
%\begin{equation}
%F(k)=\left|\int f(\mathbf{x})e^{-ik\mathbf{\hat{n}}\cdot\mathbf{x}}\mathrm{d}^3\mathbf{x}\right|^2.
%\label{formfactor}
%\end{equation}
The beam is assumed to have a simple gaussian charge distribution in the transverse $f_{\perp}(\mathbf{x}_{\perp})=(2\pi\sigma_r^2)^{-1}$exp$(-r^2/2\sigma_{r}^2)$ and longitudinal $f_z(z)=(2\pi\sigma_z^2)^{-1/2}$exp$(-z^2/2\sigma_z^2)$ dimensions. 
%The form factor is evaluated analytically with $\mathbf{\hat{n}}\cdot\mathbf{x}=r\cos\varphi\sin\theta\cos\phi+r\sin\varphi\sin\theta\sin\phi+z\cos\theta$ to obtain the coherent photon distribution for a bi-gaussian bunch:
% \begin{equation}
%\frac{\mathrm{d}^2U_C}{\mathrm{d}k \mathrm{d}\Omega}=\frac{N_e^2e^2}{4\pi^3\epsilon_0}\frac{\sin^2{\theta}\phantom{l}\mathrm{exp}\Big[-k^2(\sigma_z^2\cos^2\theta+\sigma_r^2\sin^2\theta)\Big]}{(1-\beta^2\cos^2{\theta})^2},
%\label{eq:didw}
%\end{equation}
The actual longitudinal and transverse beam sizes are $\sigma_z$ and $\sigma_r$, respectively. It is interesting to note the special case where the beam is symmetric, where $\sigma_r=\sigma_z$ and the electron beam size has no effect on the angular distribution in Eq. (\ref{eq:didw}), which becomes that of single particle emission. In that case, the emission spectrum is unpolluted by finite beam size effects and  the rms spectral width $\sqrt{\langle k^2 \rangle}=(\sqrt{2}\sigma_{z,r})^{-1}$ is directly related to the actual beam size. 

In the case of a pancake beam ($\sigma_r > \sigma_z$) considered here, however, finite beam size effects modify the spectral emission as they tend to narrow the characteristic forward hollow intensity distribution to well within the $1/\gamma$ cone. This occurs because the region for coherent emission is pushed closer to the axis as the transverse separation distance between the emitting electrons grows larger. Combined with the axial null of the single particle TR distribution kernel, this effect tends to suppress the higher frequencies that are nearest the emission axis. 
Since the width of the emission spectrum is directly related to the inverse bunch length, the suppression of higher frequency components (i.e., narrowing of the spectrum) leads to a measured bunch length value that is greater than the actual bunch length. Accurate bunch length determination requires that this effect be corrected, with a factor that may be significant depending on the beam parameters and radiation acquisition methods.

Assuming only that $\sigma_r > \sigma_z$, the exact solutions to Eq. (\ref{eq:didw}) for the spectral distribution $\mathrm{d}U_C/\mathrm{d}k$ and the total energy $U_C$ do not have a transparent, compact form. The spectrum is obtained by integration over the forward angle $0\le\theta\le\pi/2$ and is given in terms of the multivariate confluent hypergeometric functions $\Phi_1$ from \cite{GR}:
 \begin{equation}
\frac{\mathrm{d}U_C}{\mathrm{d}k}=\frac{N_e^2e^2e^{-k^2\sigma_r^2}}{2\pi^2\epsilon_0}\left[\Phi_1\left(\frac{1}{2},2,\frac{3}{2};\beta^2,k^2(\sigma_r^2-\sigma_z^2)\right)-\frac{1}{3}\Phi_1\left(\frac{3}{2},2,\frac{5}{2};\beta^2,k^2(\sigma_r^2-\sigma_z^2)\right)\right].
\label{eq:spec}
\end{equation}
The total energy is in terms of the Appell hypergeometric function $F_1$,
 \begin{equation}
U_C=\frac{N_e^2e^2}{4\pi\sqrt{\pi}\epsilon_0\sigma_r}\left[F_1\left(\frac{1}{2},2,\frac{1}{2},\frac{3}{2};\beta^2,1-\frac{\sigma_z^2}{\sigma_r^2}\right)-\frac{1}{3}F_1\left(\frac{3}{2},2,\frac{1}{2},\frac{5}{2};\beta^2,1-\frac{\sigma_z^2}{\sigma_r^2}\right)\right].
\label{eq:tot}
\end{equation}
In the limit of a strongly pancaked beam with $\sigma_r \gg \sigma_z$, the total energy emitted is $U_C\simeq N_e^2e^2\gamma/16\sqrt{\pi}\epsilon_0\sigma_r$. Note that $U_C$ is independent of $\sigma_z$ in this limiting case.

The longitudinal current profile is obtained from the width of the spectrum. The exact rms \emph{measured} bunch length $\sigma_{zm}$ is thus,
\begin{equation}
\sigma_{zm}=\frac{1}{\sqrt{2\langle k^2\rangle}}=\sigma_r\left[\frac{F_1\left(\frac{1}{2},2,\frac{1}{2},\frac{3}{2};\beta^2,1-\frac{\sigma_z^2}{\sigma_r^2}\right)-\frac{1}{3}F_1\left(\frac{3}{2},2,\frac{1}{2},\frac{5}{2};\beta^2,1-\frac{\sigma_z^2}{\sigma_r^2}\right)}{F_1\left(\frac{1}{2},2,\frac{3}{2},\frac{3}{2};\beta^2,1-\frac{\sigma_z^2}{\sigma_r^2}\right)-\frac{1}{3}F_1\left(\frac{3}{2},2,\frac{3}{2},\frac{5}{2};\beta^2,1-\frac{\sigma_z^2}{\sigma_r^2}\right)}\right]^{1/2},
\label{eq:size}
\end{equation}
It is clear that the reconstructed bunch length obtained from the CTR spectrum depends on the beam energy and the transverse size. Though a simple form of the analytic scaling is not readily apparent, Eq. (\ref{eq:size}) can be easily solved numerically as a function of $\sigma_z$ for measured values of $\sigma_r$ and $\gamma$. The the actual gaussian bunch length $\sigma_z$ can then be extracted from the measured one, $\sigma_{zm}$.

\subsection{Experimental Description and Results}
Based on the theory of "pancake-like" beams described above, an experiment was conducted at the BNL ATF to examine the consequences of large transverse size beams as it relates to CTR interferometry. 

The Brookhaven National Laboratory Accelerator Test Facility (BNL ATF) is an accelerator facility that employs a magnetic chicane to compress the electron bunches from 10ps to the sub ps level. The bunch length of such compressed bunches has been verified using interferometric techniques with coherent transition radiation (CTR) and coherent edge radiation (CER) \cite{Geloni,BoschPRSTAB}.

\begin{figure}[htb]
  \includegraphics[height=.4\textheight]{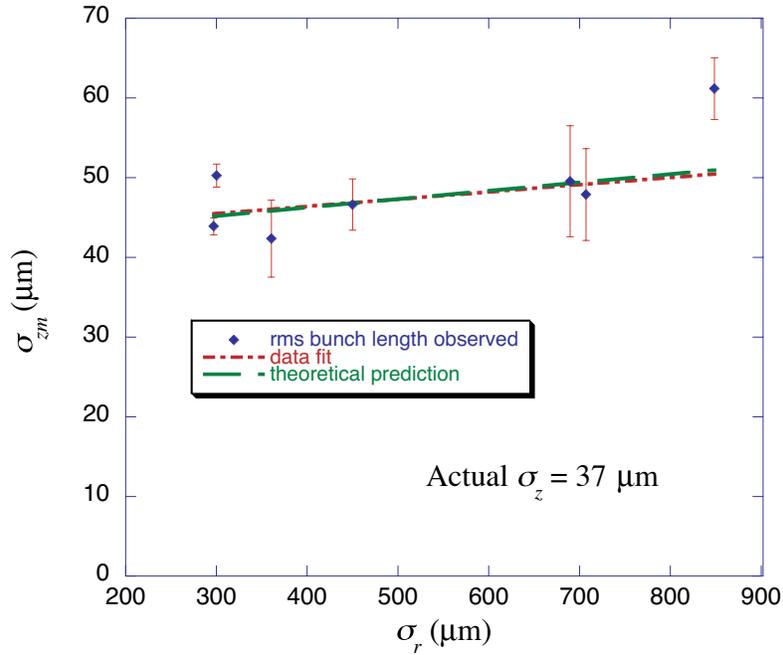}
  \caption{Comparison of bunch length determined for theoretical case (green dash-dot) and measured data (blue dots). Trendline in red dashes from least squares fit of data.}
\label{fig:data}
\end{figure}

This setup has been used for other experiments including the observation of coherent edge radiation, phase space tomography and the observation of beam breakup due to the compression process\cite{zhou}.  Details of the setup can be found in the References \cite{zhou, yakimenko}
For the CTR measurements, the electron beam is compressed in the chicane and strikes a metal mirror immediately downstream. The resultant radiation is extracted through a z-cut quartz window and passed through a Michelson-type interferometer which can accommodate wavelengths up to 1.5~mm. 
The autocorrelated signal is detected on a single Golay cell detector and the output is measured on an oscilloscope.
The electron beam parameters for these measurements are an energy of 61~MeV and bunch charge of 350~pC.

The data from the CTR measurements is presented in Figure~\ref{fig:data}. The CTR autocorrelation traces were averaged over 5 shots per position and a bi-gaussian fitting method was used to determine the bunch length \cite{murokh}. The blue dots (with red vertical error bars) show the calculated bunch length for various rms transverse beam sizes  ranging from 300~$\mu$m to 850~$\mu$m.
The dashed red line is a best fit curve for the measured data. 
The dashed green line is the theoretical model of Equation~\ref{eq:size} for an actual bunch length of $\sigma_z$=37~$\mu$m.
There is strong agreement between the analytical curve and the best fit trendline through the data indicating that the predicted effect of large transverse beam size is a real effect that must be considered.

Currently, work is being performed to develop a simpler analytical correction factor from Eq. (\ref{eq:size}) that can be used to account for real beam sizes when reconstructing bunch length using CTR based interferometric methods.

\section{Conclusions}
  
Interferometric methods on beam-based radiation are commonly used to determine the bunch length of compressed beams. 
However, the experimental results based on the autocorrelation of CTR signals must be corrected for beams with large transverse-to-longitudinal aspect ratios. 
The CTR emitted by these type of beams inherently suppresses large frequency components yielding a reconstructed bunch length greater than the actual bunch length. 

This effect is correctable by adjusting the analysis to take into account the apparent pulse broadening due to the transverse contribution to the autocorrelation signal.
This method was examined at the BNL ATF and the results compare well with the analytical model. Further studies will address other aspects of more generalized beam shapes and other beam-based coherent radiation interferometry (such as diffraction or edge radiation). 
The results are important for future light sources and advanced accelerator facilities where accurate monitoring of the bunch compression process is imperative for experimental operations \cite{akre}.

\begin{theacknowledgments}
Work supported by DOE Grants Nos. DE-FG02-07ER46272 and DE-FG03-92ER40693, and ONR Grant No. N00014-06-1-0925.
\end{theacknowledgments}

\bibliographystyle{aipproc}   % if natbib is available

\bibliography{pancake_01}

\end{document}